\documentclass[aps,twocolumn,floatfix]{revtex4}
\usepackage{amsmath}
\usepackage{graphicx}
\usepackage{booktabs}
\usepackage{textcomp}
\usepackage{dcolumn}
\usepackage{sidecap}
\usepackage{subfigure}

\begin{document}
\title{Penning traps with unitary architecture for storage of highly charged ions}
\author{Joseph N Tan$^1$, Samuel M Brewer$^2$, and Nicholas D Guise$^1$ }
\affiliation{$^1$ National Institute of Standards and Technology, 100 Bureau Drive,
Gaithersburg, Maryland 20899-8422, USA}
\affiliation{$^2$ University of Maryland, College Park, Maryland 20742, USA}
\date{\today}

\begin{abstract}
Penning traps are made extremely compact by embedding rare-earth permanent magnets in the electrode structure.   Axially-oriented NdFeB magnets are used in unitary architectures that couple the electric and magnetic components into an integrated structure.  We have constructed a two-magnet Penning trap with radial access to enable the use of laser or atomic beams, as well as the collection of light.  An experimental apparatus equipped with ion optics is installed at the NIST electron beam ion trap (EBIT) facility, constrained to fit within 1 meter at the end of a horizontal beamline for transporting highly charged ions.  Highly charged ions of neon and argon, extracted with initial energies up to 4000 eV per unit charge, are captured and stored to study the confinement properties of a one-magnet trap and a two-magnet trap. Design considerations and some test results are discussed.

[http://link.aip.org/link/?RSI/83/023103] Copyright 2012 American Institute of Physics 
\end{abstract}

\maketitle

\section{Introduction}


The Penning trap \cite{Penning1936} is widely used to confine charged particles in an evacuated region of space through the use of static electric and magnetic fields.    A versatile tool for enabling isolation and manipulation of charged particles in a well-controlled environment, it finds applications in various disciplines, including physics of non-neutral plasmas, mass spectrometry, biomolecular chemistry, precision spectroscopy, antimatter science, quantum information and fundamental metrology.\cite{Geonium1986,UWalpha1987,JJB1994NIST,BeCrystal1998,FTICR1998,Kluge2003,Surko2004,GGHbar,GGMass,eQbit2008,eQbitJJB2009,eQbit2010GG,GGalpha2010} An iconic example is the creation of an artificial atom called ``geonium,"\cite{Dehmelt1990} consisting of a single elementary particle isolated in a Penning trap;\cite{Geonium1986} the most precise value of the fine structure constant $\alpha$ is determined from the spin flip and cyclotron oscillation of one electron in such a system.\cite{UWalpha1987,GGalpha2010,CODATA2010}

A Penning trap is made of a stack of cylindrically-symmetric electrodes biased to generate a restoring force along the trap axis.  For radial confinement this electrode stack and its vacuum envelope are typically inserted into a coaxial solenoid which can generate a strong magnetic field ($>> 1$ Tesla) along the common symmetry axis. A superconductive magnet, such as used in nuclear magnetic resonance (NMR) experiments, can provide very uniform magnetic fields greater than 5 Tesla.  This classic architecture separates the electrode structure and the source of magnetic field, allowing the two subsystems to be independently removed, modified or adjusted when necessary.  For applications demanding high precision, a superconductive solenoid can be designed to provide (1) field homogeneity as high as 1 part in  $10^8$ over 1 cm$^3$, and (2) self-shielding via flux conservation to screen out ambient field fluctuations.\cite{S4shield}  On the other hand, a superconductive magnet is costly and may not be entirely advantageous for some applications. NMR magnets can occupy as much as $\approx\! 1$ m$^3$ of space and require some form of refrigeration. This is not compatible with small instrument development, nor with facilities or missions that have severe space constraints.   

Rare-earth permanent magnets with high remnant fields have enabled the construction of a variety of compact structures with strong magnetic fields.\cite{Halbach1980}  Compact Penning traps have been developed which utilize arrays of radially-oriented magnets (wedges) to replace a solenoidal magnet.\cite{Gomer1995, Dunning2002} In this work we introduce a simplified, unitary architecture for compact Penning traps and discuss experiments using such traps for storage of highly charged ions extracted from an EBIT ion source.  In this unitary architecture, the rare-earth magnets that generate the magnetic field are naturally integrated within the electrode structure that generates the electric field.  Details regarding the slowing and capture of highly charged ions from the NIST EBIT will be presented in a separate publication.  Here we focus upon the design and the observed properties of unitary Penning traps with embedded magnets. Section \ref{sec:DesignConstruct} presents two architectures utilizing axially-oriented rare-earth magnets.   In Section \ref{sec:Observations} we describe an experimental apparatus deployed at the NIST EBIT facility and present the first results from tests using stored highly charged ions.  


\section{Design and Construction}
\label{sec:DesignConstruct}

We are interested in isolating ions extracted from the NIST EBIT for spectroscopic experiments.\cite{1eRydberg2008,1eRydberg2009,HCI2010Tan} Space constraints preclude the use of a multi-Tesla Penning trap\cite{Retrap2005}. Since compact Penning traps using rare-earth magnets occupy less than 1000 cm$^3$ of space, they are an attractive alternative. Such compact Penning traps have been used to study light ions in a liquid-nitrogen cooled apparatus\cite{Gomer1995} and to store molecular anions in a room-temperature apparatus\cite{Dunning2002}. For our planned experiments, however, the trap must be operable at the high voltages necessary to capture highly charged ions.  Very low background gas pressure must be attainable at room temperature, and access must be provided for laser or atomic beams to interact with the stored ions. Such constraints and requirements, together with known conditions favorable for ion confinement, have led us to explore non-traditional architectures which could help simplify construction. The simplest Penning trap using one permanent magnet is discussed in Sec. \ref{sec:OneMag}, illustrating the basic features of a unitary architecture. This is followed by a more intricate design involving two rare-earth permanent magnets (Sec. \ref{sec:TwoMag}) to provide better magnetic field homogeneity and radial access for laser or atomic beams.

\subsection{One-magnet Penning trap}
\label{sec:OneMag}

\begin{figure}[]
\includegraphics[angle=0,width=0.47\textwidth]{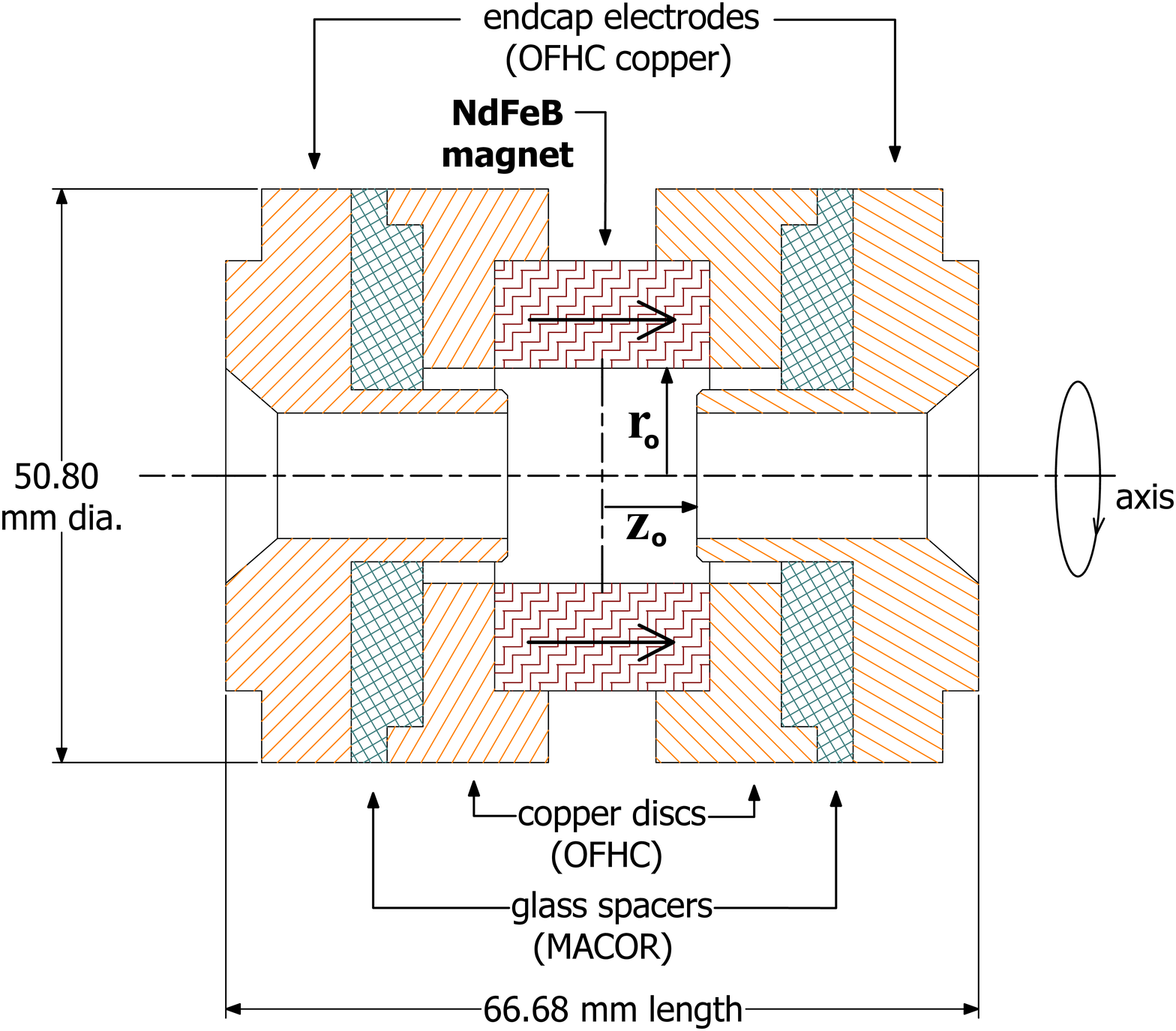}
\caption{\label{fig:1magtrapcad} Schematic of the one magnet Penning trap; the symmetry axis is horizontal. The NdFeB magnet is the central ring electrode, with its axial magnetization $\bf{M}$ indicated by the arrows.  Characteristic dimensions of the trapping volume: $r_o = 9.525$ mm is the inner radius of the NdFeB magnet; $z_o = \pm 8.385$ mm is the distance from the midplane to one of the endcaps.}
\end{figure}

A simple design for a Penning trap using an embedded permanent magnet is shown in the schematic diagram Fig. \ref{fig:1magtrapcad}. To allow ion passage, open endcap electrodes (OFHC copper) are used, separated from the central ``ring" electrode by insulating spacers made of machinable glass ceramic(MACOR\cite{disclaimer}). The ring electrode is a composite structure consisting of an axially-magnetized, annular neodymium magnet sandwiched tightly between two annulated copper discs. The annulated copper discs on each side are used to align the magnet with the axis of the electrode stack, and have tapped holes for attaching wires. Rare-earth magnets are fabricated from compounds containing an element in the Lanthanide series, such as praseodymium ($^{59}$Pr), neodymium ($^{60}$Nd) or samarium ($^{62}$Sm). The neodymium magnets, chemically denoted by Nd$_2$Fe$_{14}$B or NdFeB, are chosen for our work because they are readily available in various grades and shapes with triple-layer plating to protect the magnets from corrosion; the first layer is nickel, followed by copper and finally nickel again. Hence, an electrical potential can be applied to the full surface of the composite ring electrode via one of the press-fitted copper discs. 

The integration of the rare-earth magnets into the electrode structure for providing the trapping fields is an essential feature of a unitary architecture.  This coupling of the magnetic and electric components is fully exploited in Section \ref{sec:TwoMag} wherein the trap electrodes are made of iron to yoke the fields emanating from the rare-earth magnets. The dual role played by the NdFeB magnets and electrodes reduces the number of trap components and the overall size of the ion trap while still providing a useful magnetic field of $\approx $ 0.317(1) T in the trapping region. The NdFeB magnet (N42 grade) has dimensions of 19.05 mm inner diameter, 38.10 mm outer diameter, and 19.05 mm length. The trapping region has characteristic dimensions of $r_o = 9.525$ mm and $z_o = 8.385$ mm.  The overall assembly diameter and length of the trap are $50.8$ mm and $66.7$ mm, respectively.

\begin{figure*}[]
\includegraphics[angle=0,width=6.45in]{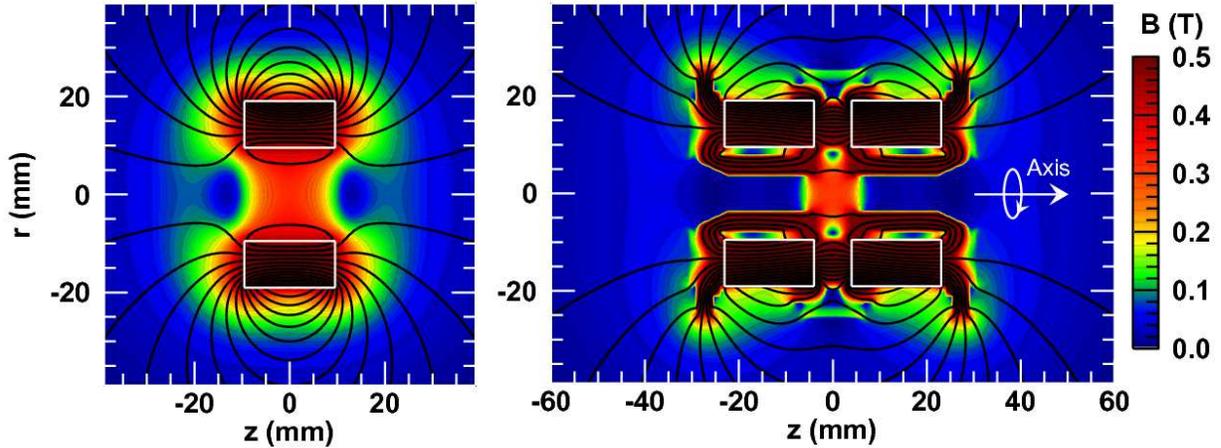}
\caption{\label{fig:BfieldMap} (Left) Magnetic flux density of a NdFeB magnet (N42 grade). The axis of the magnet is horizontal. Magnet size: 19.05 mm inner diameter, 38.10 mm outer diameter, and 19.05 mm length. (Right)Magnetic flux density for the two-magnet Penning trap shown in Figure~\ref{fig:2magtrapcad}, which uses N40UH grade NdFeB magnets. The axis of rotational symmetry is horizontal.  Colors indicate equal B-field contours, with the B-field inside the magnet arbitrarily pegged at 0.5 Tesla to highlight the trapping region. The flux density inside a magnet is as high as $\approx\! 1.2$ Tesla.}
\end{figure*}

Figure \ref{fig:BfieldMap} shows the magnetic flux density (B-field) produced by a single NdFeB magnet, computed using QuickField\cite{disclaimer}, a commercial software which implements a well-known finite element method (FEM)\cite{Campbell1994}. The magnetic B-field ($\bf{B} = \mu_o\bf{H} + \bf{M}$) has a saddle-point at the center.  The magnet field is 0.317 Tesla at the trap center, growing radially to 0.381 T at the inner wall of the magnet.  On axis, the field drops to zero at a distance of 12.94 mm from the center, reversing direction and reaching a local maximum of 0.089 T at 20.75 mm  before attenuating with increasing distance.  This field pattern is identical to that of a model based on a pair of solenoids at the inner and outer walls of the magnet, carrying counter-rotating currents.\cite{Panofsky1962} 


Near the center of the trap the ion motions are well approximated by calculations for an ideal Penning trap\cite{Geonium1986} wherein a uniform magnetic field is superimposed upon a quadrupole potential.  In a plane perpendicular to a uniform magnetic field, a charged particle will undergo circular motion with a frequency, commonly called the cyclotron frequency, given by
\begin{equation}
\label{cycf}
\omega_{c} = \frac{Qe|B|}{m} ,
\end{equation}
where $Qe$ is the charge of the particle, $|B|$ is the magnitude of the magnetic field, and $m$ is the mass of the particle.  For a bare Ne nucleus in a magnetic field 
of $\approx$ 0.3 T, the cyclotron frequency $\omega_{c}/2 \pi \approx$ 2.3 MHz.  Along the trap axis of symmetry (z-axis), the center of cyclotron orbit bounces between the endcaps, which are positively biased relative to the ring electrode; to lowest order, the electric field is given by a quadrupole potential
\begin{equation}
\label{pennV}
V(r,z) = \lambda V_0 \frac{z^2 - r^2/2}{2d^2} + \rm{constant}  .
\end{equation}
The field coordinates $z$ and $r$ are defined from the center of the trap and $d$ is determined by trap dimensions 
\begin{equation}
\label{d}
d^2 = \frac{1}{2} (z_0^2 + r_0^2/2) 
\end{equation}
where $r_0$ is the inner radius of the ring electrode, and $z_0$ is the distance of reentrant endcaps from the midplane.  
The coefficient $V_0$ is the potential difference between the endcaps and the central ring electrode; it is used interchangeably with $\Delta V$ in other sections.  The dimensionless parameter $\lambda$ (sometimes denoted C$_2$)is a geometrical factor of order unity; if the electrode surfaces near the center closely approximate hyperbola of revolution, then $\lambda \approx 1 - \epsilon$ with $0 < \epsilon << 1$. Small amplitude motion along the trap axis is described by a simple harmonic oscillator with frequency
\begin{equation}
\label{axf}
\omega_z^2 = \lambda \frac{QeV_0}{md^2} .
\end{equation}
For the one-magnet trap in Fig. \ref{fig:1magtrapcad} with $V_0=10$ V, an axially-bound $^{20}$Ne$^{10+}$ ion oscillates with $\omega_z / 2\pi \approx$ 414.8 kHz. For this trap, $\lambda = 0.814$.

In contrast to the stable axial oscillation, transverse motion is localized only if dynamical equilibrium is possible. The axial restoring force provided by the quadrupole electric field is accompanied by an outward radial force on the ion, pulling the ion towards the ring electrode. This tendency to leave the trap radially (i.e., to roll off the saddle point of Eq.\ref{pennV}) must be balanced by the Lorentz force due to the magnetic field if the ion is to remain trapped. Consequently, a third component of the ion motion, the magnetron motion, arises from the $\mathbf{E} \times \mathbf{B}$ interaction\cite{Geonium1986}. In equilibrium the cyclotron orbit drifts slowly around the center of the trap, as illustrated in Sec.\ref{sec:TwoMag}, with a magnetron frequency given by
\begin{equation}
\label{magf}
\omega_m = \frac{1}{2}\left[\omega_c - \sqrt{\omega_c^2 - 2\omega_z^2}\right] .
\end{equation}
Considering a bare Ne nucleus with the trapping parameters described above, $\omega_m /2\pi \approx$ 38.03 kHz.  
 
The condition for equilibrium is more restrictive in weaker magnetic field. As discussed in Ref.\cite{Geonium1986}, magnetron motion requires
\begin{equation}
\label{stabcond}
\omega_c^2 - 2\omega_z^2 > 0.
\end{equation}
Hence, the existence of localized motions is determined by the trap geometry, applied fields and ion properties (charge and mass). 
In an ideal Penning trap, if this condition is satisfied at one location, then it holds over the entire volume bounded by the electrodes.  In a one-magnet Penning trap the allowed region could be reduced by its magnetic field gradient; useful axial well depth is restricted to a narrower range for a one-magnet trap than for a high-field solenoid system. Nevertheless, the trap illustrated in Fig. \ref{fig:1magtrapcad} proves sufficient for capture of highly charged ions from an EBIT.  We observe ion storage times of order 1 second, limited primarily by background gas collisions(Sec. \ref{sec:Observations}).


\subsection{Two-magnet Penning trap}
\label{sec:TwoMag}

The one-magnet Penning trap allows straightforward assembly and operation at high voltages, but it has some drawbacks.  The NdFeB magnet blocks laser or atomic beams propagating in the midplane, obstructing access to the stored ions. Since the entire inner surface of the NdFeB magnet is in the line of sight from the trap center, magnet imperfections could degrade ion confinement stability. Furthermore, the uniformity of its magnetic field is not optimized.  A possible solution is to construct a compact Penning trap using the classic architecture but with the solenoid magnet replaced by two radially-magnetized rings, each formed from 8 NdFeB wedges.\cite{Gomer1995} However, the construction of such magnet arrays requires considerable effort and care to ensure good alignment and axial symmetry. In addition, adhesives used in bonding the wedges(e.g., epoxies) generally degrade in high temperature bake-outs commonly used in improving base pressure for room-temperature applications.


In this section, we present a simpler alternative: an architecture employing two identical NdFeB magnets ($\bf{M}$ parallel to their aligned symmetry axes) and electrodes made of soft iron in an integrated structure for producing the magnetic field.  The design work is guided by FEM simulations of magnetic structures using QuickField. The precision of the algorithm depends upon the coarseness of the grid and the accuracy of the magnetization (B-H) curves (for materials such as NdFeB\cite{MMPA100}, soft iron, etc.). 

\begin{figure}[]
\includegraphics[angle=0,width=0.48\textwidth]{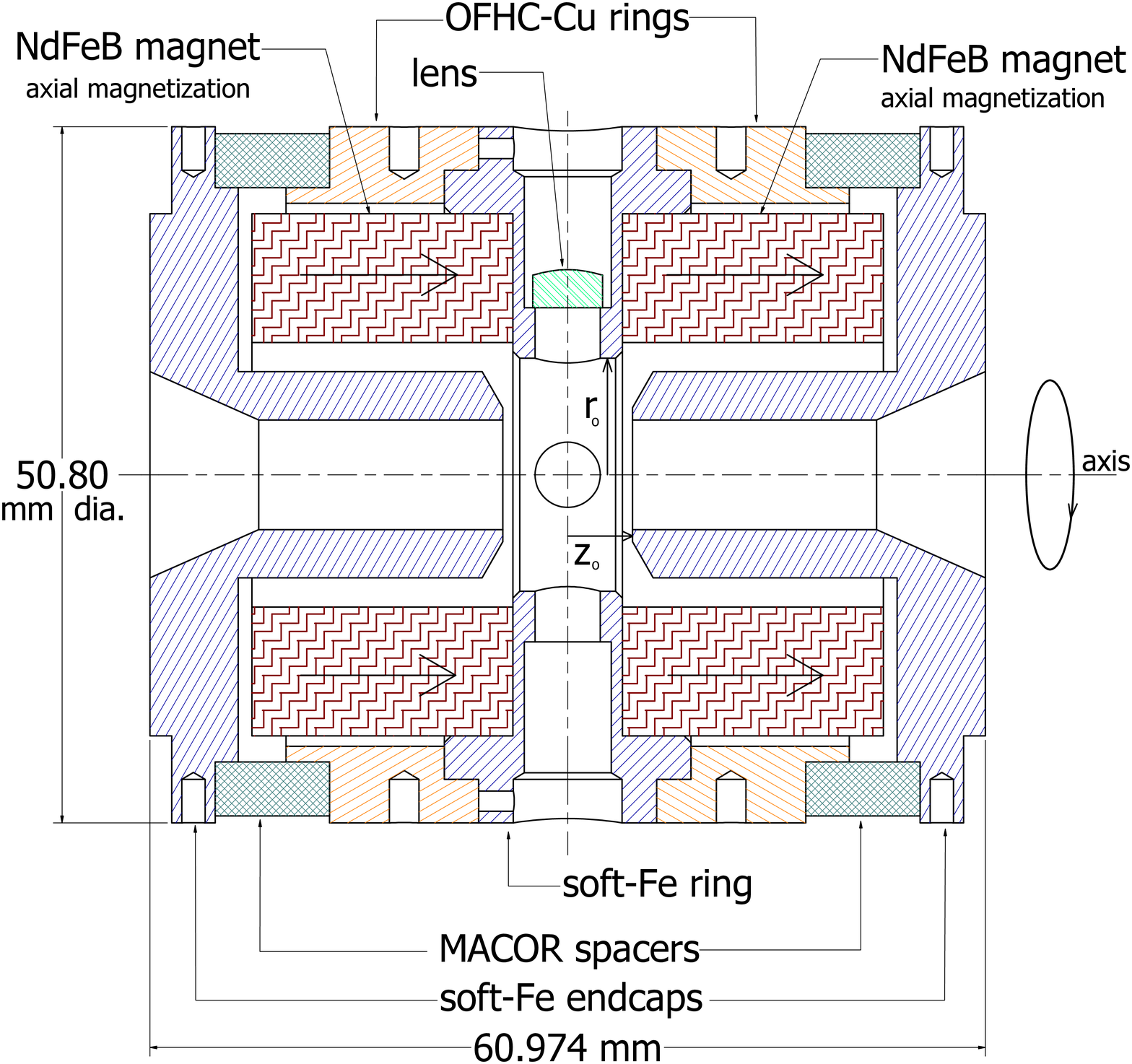}
\caption{\label{fig:2magtrapcad} Diagram of a Penning trap with two embedded NdFeB magnets; the symmetry axis is horizontal. The two magnets are seated tightly on opposite sides of the iron ring electrode with inner radius $\rm{r}_o = 8.500$ mm.  Holes in this ring allow beam access.  The iron endcaps have reentrant tubes extending into the NdFeB magnets and ending at $\rm{z}_o = \pm 4.736$ mm from the midplane.  The outer diameter is 5.08 cm, and the overall length is 6.10 cm.}
\end{figure}

Figure \ref{fig:2magtrapcad} shows a diagram of the two-magnet Penning trap.  The two open-access endcaps and the central ring electrode are made of electrical iron, with inner surfaces polished to mirror-like finish.  To enable radial beam access to the stored ions, the central ring electrode has four equidistant holes, along two orthogonal directions.  The top hole in the vertical direction houses an aspheric lens for light collection. The characteristic trap dimensions are $r_o = 8.500$ mm (inner radius of central ring) and $z_o = 4.736$ mm (distance from the midplane to an endcap).  The overall assembly diameter and length of the trap are 50.8 mm and 60.974 mm, respectively. Near the center of the trap, a bare Ne nucleus bound in a $V_0=10$ V well would oscillate with axial frequency $\omega_z / 2\pi \approx$ 597.04 kHz, corresponding to $\lambda = 0.854$ in Eq.\ref{pennV}.

 The rare-earth magnets are of the same dimensions as the one used in the one-magnet design (Sec. \ref{sec:OneMag}), but are manufactured for continuous operation at temperatures up to 453 K or 180 $^{\rm{o}}$C (N40UH grade).  N40UH grade magnets are chosen to withstand higher bakeout temperatures without loss of magnetization.  The two magnets are oriented to have parallel magnetization.  One at a time, the magnets are fitted tightly onto the ring electrode; since there is strong magnetic attraction between the magnet and the iron ring, an installation device with a threaded rod is used to control the docking process. A similar installation procedure is used to insert the reentrant endcaps into the magnets. Each endcap is aligned with the central ring electrode directly, using a copper ring and MACOR spacer in series.  
 
 This two-magnet architecture with soft-iron electrodes has several useful features.  First, the alignment of this trap is better than the one-magnet design because it relies only upon fitting precision-machined components. Since the NdFeB magnets are secluded behind the reentrant endcaps, magnet imperfections which could degrade ion confinement are attenuated. The electrical-iron endcaps help to reduce the fringing magnetic field, and their reentrant shape yokes the field towards the trap center.  By redirecting the magnetic flux around the magnets to close the magnetic circuit via the trapping region, the soft-iron endcaps largely account for the strength and uniformity of the magnetic field near the trap center.   

\begin{figure}[]
\includegraphics[angle=0,width=0.49\textwidth]{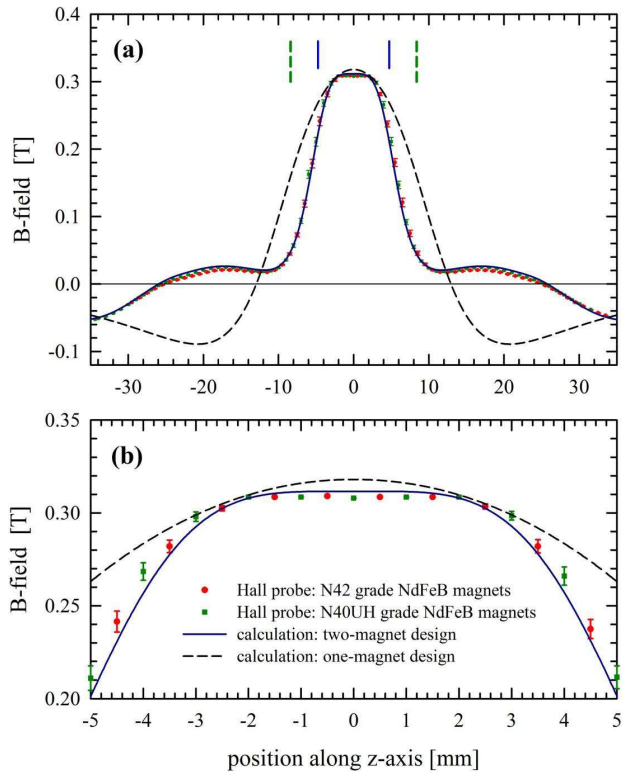}
\caption{\label{fig: 2mbfield} (a)Comparison of the on-axis magnetic field calculated for the one-magnet trap (dash line) and the two-magnet trap (solid line); (b)magnification for finer comparison of the homogeneity within the trapping region.    The reentrant edges of the trap endcaps are indicated by a pair of vertical lines; green dashed line for the one-magnet trap and blue solid line for the two-magnet trap. Hall probe measurements for the two magnet trap are also plotted: circle for N42 grade neomagnets, and square for N40UH neomagnets.}
\end{figure}

A comparison of the calculated and measured axial magnetic field is shown in Fig. \ref{fig: 2mbfield}.   An axial Hall probe is used to measure the field along the trap axis. The Hall sensor has dimensions 1.9 mm by 2.3 mm wide, and $\approx\! 0.5$ mm thick; the overall instrument accuracy is about 2 \%, but the resolution is $\approx\! 1$ part in 300 in the range of the measurements reported here. Two grades of NdFeB magnets were used: N42 (red circle) and N40UH (green square). Figure \ref{fig: 2mbfield}(a) compares measurements with calculation (solid line) for the on-axis magnetic field.  An axi-symmetric model is used, neglecting perturbations due to the 4 holes in the ring electrode.  When magnetization (BH) curve parameters are optimized within typical manufacturing tolerance (5 \% to 8 \%),\cite{MMPA100} the model calculation fits well to data over the range $-3$ cm $< z < 3$ cm. Figure \ref{fig: 2mbfield}(b) gives a detailed view of the trapping region of the two-magnet Penning trap. The error bars represent $1\sigma$ uncertainty, combining in quadrature the probe resolution with the uncertainty $\delta z (dB/dz)$ due to the field gradient. The scatter of measured B-field values in the region $\pm 2$ mm from the trap center has a standard deviation that is smaller than the sensor resolution.  This suggests that the homogeneity is about 1 part in 300 in the optimized region; higher uniformity should be obtainable using small shim coils.  Compared with a one-magnet trap (dash line), the magnetic field homogeneity in the two-magnet trap is better within $\pm 2$ mm from the trap center.

The electrostatic field near the center of the trap is approximately that of a quadrupole potential.  Notwithstanding holes for admitting ions or radial beams, the electrode surfaces facing the trap center are chosen to help minimize deviations from a quadrupole potential.\cite{Beaty1987} Different geometries were studied numerically using a Boundary Element Method (BEM)\cite{BEM1976Read} originally developed for calculating properties of electrostatic lenses. In the two-magnet trap the reentrant endcaps are positioned close to the trap center in order to make the magnetic field more uniform, causing the electric field due to the endcaps to penetrate the trap center. As a result, the axial trapping well has a depth (maximum-to-minimum potential difference) equal to 38.8 \% of $\Delta \rm{V} \equiv V_0 $, the applied voltage.

\begin{figure}[]
\begin{center}
\includegraphics[angle=0,width=0.43\textwidth]{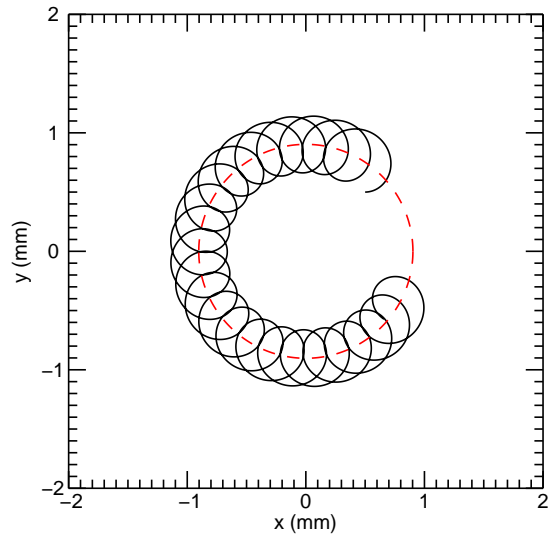}
\caption{\label{fig: xysim} Circular motions of a Ne$^{10+}$ ion in the midplane of the two-magnet Penning trap.  The initial condition is chosen to illustrate the fast cyclotron motion undergoing a counter-clockwise magnetron drift (dashed circle) around the center of the trap.}
\end{center}
\end{figure}

Having obtained the electric and magnetic fields of the trap, the motions of an ion are computed by integrating the equations of motion using standard Runge-Kutta techniques. An example of an ion trajectory is shown in Fig. \ref{fig: xysim}. Similar computations were undertaken to investigate the conditions under which energetic ions extracted from an EBIT ion source can be slowed and captured.

\section{Experiments}
\label{sec:Observations}


\begin{figure*}[t]
\begin{center}
\includegraphics[angle=0,width=6.45in]{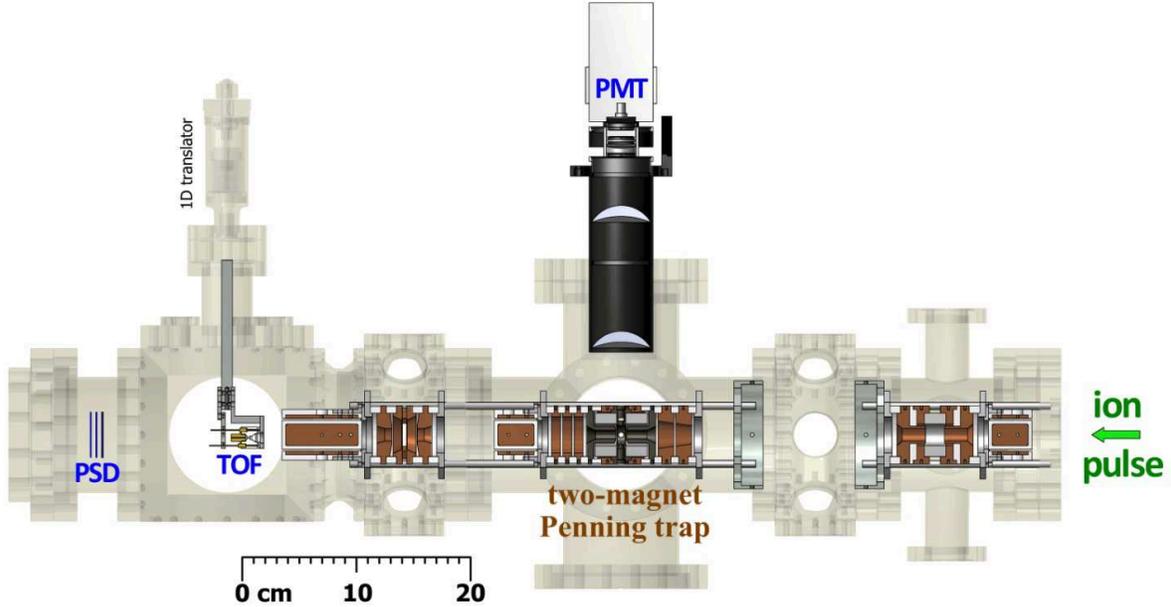}
\caption{\label{fig:Apparatus} Simplified diagram of the experimental apparatus. The two-magnet Penning trap is centered on the six-way cross. An ion pulse is steered and focused by orthogonal plates and an Einzel lens (right).  Ions are counted using the time-of-flight (TOF) detector or the position-sensitive detector (PSD).}
\end{center}
\end{figure*}

Fig. \ref{fig:Apparatus} shows a simplified diagram of a room-temperature apparatus with the two-magnet Penning trap (centered on the six-way cross), ion optics, and several detectors.   The vacuum chamber (illustrated in very light shades) is built from readily-available commercial components, evacuated using turbo-molecular and ion pumps (not shown).  Ions entering the apparatus are first steered by orthogonal pairs of deflectors (rightmost) and then focused with an Einzel lens.  As the ions approach the Penning trap in the six-way cross, they are slowed down by the electric field from a pair of rings with tapered inner surfaces.  The two-magnet Penning trap is oriented so that a pair of holes in the ring electrode is aligned with the center of a reentrant window at the top of the six-way cross; a lens system installed on top of this cross collects light emitted by the stored ions onto a photomultiplier.  Ions can be counted by ejecting them to a time-of-flight (TOF) micro-channel plate (MCP) centered on the six-way cube; alternatively, a position-sensitive MCP detector (leftmost) can be used if the TOF detector is retracted with a translator.  This experimental set-up is constrained to fit within $\approx\! 1 \rm{m}^3$ of vacant space in the ion extraction area of the NIST EBIT.

The ion source is an electron beam ion trap (EBIT) which produces highly charged ions by electron impact ionization of an injected gas.  The NIST EBIT is equipped with a beamline for extracting the highly charged ions in pulses, analyzing the charge states, and transporting them to the user area.\cite{RatliffEBITBeamline,RatliffRobertsEBIT2001} For the work reported here, neutral gas of neon or argon is injected into the EBIT, which is operated with an electron beam energy as high as 4000 eV to produce various charge states of interest. Extracted ions have an energy of $E_{ion} = Q U_{e-beam}$, where $Q$ is the ion charge and $U_{e-beam}$ is the electron beam energy. A simple diagram illustrating the transport, charge state selection by an analyzing magnet, and injection of extracted ions into a Penning trap is provided in Ref.\cite{HCI2010Tan}.  

\begin{figure*}[t]
\begin{center}
\includegraphics[angle=0,width=5.75in]{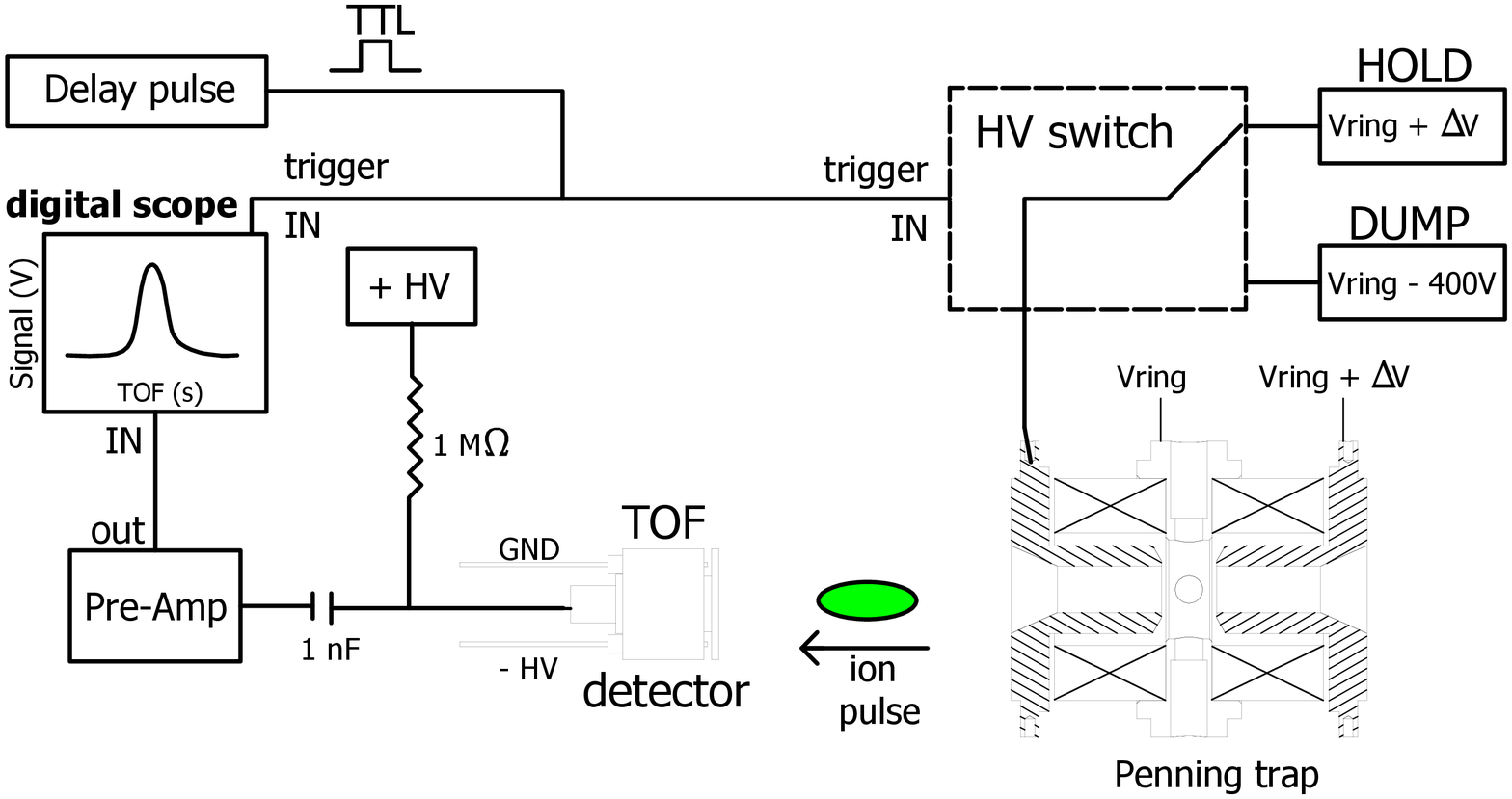}
\caption{\label{fig:DetectScheme} Diagram for ion detection scheme. A TTL pulse triggers a high-voltage switch to eject stored ions, and simultaneously triggers a digital oscilloscope to begin data acquisition of the TOF detector signal.}
\end{center}
\end{figure*}

We have recently demonstrated capture of highly charged ions from the EBIT using both the one-magnet Penning trap in Fig.\ref{fig:1magtrapcad} and two-magnet Penning trap in Fig.\ref{fig:2magtrapcad}. A detailed discussion of the transport, slowing and capture of ions with extraction energies of $E_{ion} \approx 4000Q$ eV will be presented elsewhere. In brief, an ion pulse is admitted into the Penning trap by momentarily lowering the potential on the entrance endcap during its transit from the ion source. At an optimal time corresponding to the arrival of the ions in the Penning trap the potential of this endcap is raised to close the trap. To effectively slow ions, the Penning trap is floated to a high voltage that matches the EBIT ion beam energy.  In these early tests, captured ions included Ne XI, Ne X, Ne IX, as well as Ar XVII, Ar XVI, Ar XV and Ar XIV.  In the discussion below, we focus on bare neon nuclei (i.e., Ne$^{10+}$ or Ne XI).

To evaluate the trap operation, the captured ions are kept in the trap for a particular ``storage time,'' and then ejected through the hole in the exit endcap towards a micro-channel plate detector(see Fig. \ref{fig:Apparatus}). Upon hitting the TOF detector, the ion pulse is converted by the micro-channel plate into an electrical pulse with a gain in the range from $10^5$ to $10^6$. 

The timing and detection scheme is shown in Fig. \ref{fig:DetectScheme}. The electron pulse at the output of the detector is capacitively coupled to a fast pre-amplifier which converts it into a voltage for digital acquisition. This signal is recorded by a digital oscilloscope that is triggered to start data acquisition at the same time when the trap is triggered to release the stored ions. The trigger pulse from a gate/delay generator is delayed from the instant of ion capture by the desired storage time.

\begin{figure*}[t]
\begin{center}
\includegraphics[angle=0,width=5.70in]{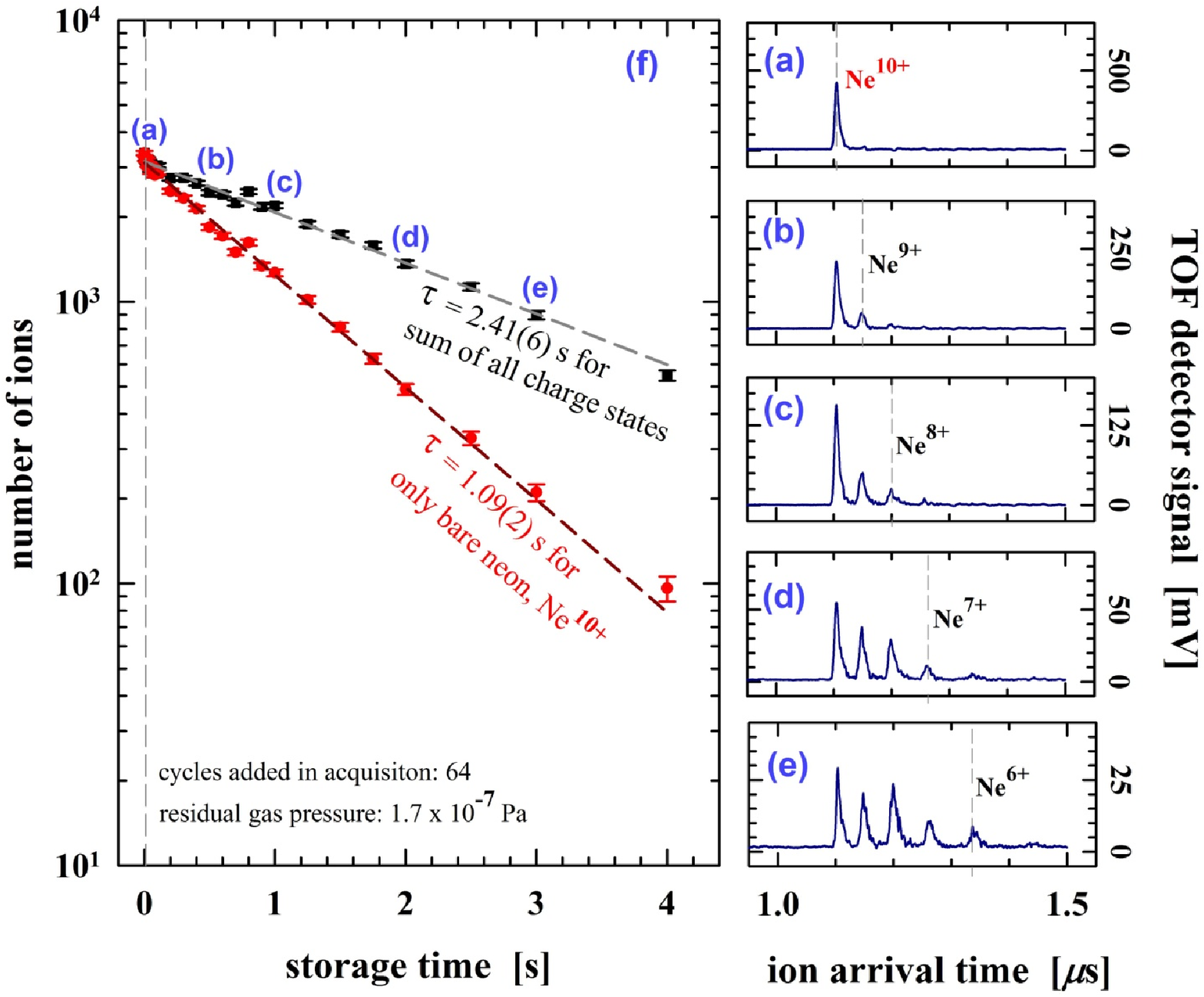}
\caption{\label{fig:TOF_IonLossDecayTime}  Storage of highly charged ions in a two-magnet Penning trap. (Left) Number of ions detected on the fast MCP as a function of storage time in the two-magnet Penning trap with $\Delta V = 10$ V applied between the ring and endcap electrodes. (Right) Output of the TOF detector versus arrival time, sampled for representative storage times:(a)10 ms (b)0.5 s (c)1 s (d)2 s (e)3 s.  The detector signal scale is magnified by $\approx\! 2\times$ stepwise from (a) through (e). The TOF peak for each charge state is converted to ion counts for (f).  The number of ions decays exponentially as a function of the storage time, as illustrated in (f). Error bars represent $1 \sigma$ uncertainty.}
\end{center}
\end{figure*}


  The TOF detector signals for some representative storage times are shown in Figure \ref{fig:TOF_IonLossDecayTime} (right side).  For storage times much shorter than a second, the TOF detector signal has only one strong peak, corresponding to Ne$^{10+}$, as illustrated in Fig.\ref{fig:TOF_IonLossDecayTime}(a).  After a storage time of about half a second (b), a second peak is clearly observable, with its arrival time corresponding to Ne$^{9+}$.  With increasing storage time from (c) through (e), more TOF signal peaks are observed.  The production of lower charge states is due to electron capture from background gas atoms in charge exchange collisions.  The vacuum chamber pressure was about $1.7 \times 10^{-7} $ Pa.
 
 The overall TOF signal size decreases with storage time. The vertical scale is magnified in going from Figure \ref{fig:TOF_IonLossDecayTime}(a) through (e), by a factor of 2 each time.  This pattern suggests that the ion cloud is expanding radially, thus reducing the number of ions that can pass through the aperture (8 mm diameter) of the open-access endcap when ejected.\cite{Oxley2004}  It is possible, particularly for long storage times, that some ions may have already collided with the ring electrode by the time of ejection.
 
The TOF detector signal can be converted to ion number since the charge states and signal gain calibration are known.  With the fast response time of the TOF detector (8 mm active diameter, $\sim\! 0.5$ ns rise time) the peaks corresponding to the arrival of different charge states are well resolved, allowing the number of ions to be determined with good accuracy.  

We observe exponential decay in the ion count as a function of the storage time, as illustrated in Figure \ref{fig:TOF_IonLossDecayTime}(f). The logarithmic-linear plot shows a good fit to an exponential function for the number of detected bare Ne ions, as well as for the sum of all detected ions including charge exchange products. The decay time constants (e$^{-1}$ time) are 1.09(2) s for the detected bare Ne nuclei, and 2.41(6) s for the sum over charge states. The decay rate for the sum of all detected ions is presumably due to the slow expansion of the ion cloud as energy is removed due to collisions with the background gas.\cite{BufferGas1968,IonLz_Wineland1985,CoolingIons1995} To the extent that the detected charge states have roughly the same detection efficiency, the decay rate for the sum of charge states provides an estimate of the expansion rate.  Subtracting the expansion rate from the decay rate for bare Ne nuclei, one obtains an estimate of the charge exchange rate of conversion to lower charge states; this gives a time constant of $\approx\! 2.0$ s for the charge-exchange loss of Ne XI ions stored in a trap with background gas pressure of $1.7 \times 10^{-7}$ Pa.

\begin{figure}[]
\begin{center}
\includegraphics[angle=0,width=0.42\textwidth]{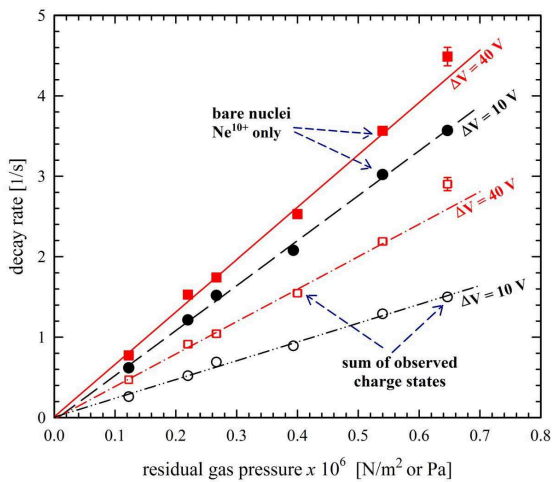}
\caption{\label{fig:DecayPressureDependence} Dependence of observed ion number decay rate upon background gas pressure, for two potential well depths: $\Delta V = 10$ V (circle) and $\Delta V = 40$ V (square). Filled symbols are for Ne$^{10+}$ only. Unfilled symbols are summed over charge states. Error bars represent $1 \sigma$ uncertainty. }
\end{center}
\end{figure}

Collisions with the residual background gas is a dominant factor for the retention of the captured ions in this room-temperature trap. The background gas cools the axial motion. However, as discussed in Sec. (\ref{sec:OneMag}), the magnetron drift that keeps an ion on the potential hill is metastable; background gas collisions (drag) tend to increase the magnetron radius until the ion hits the ring electrode.\cite{CoolingIons1995} We measured the decay rates of stored ions for various background gas pressures.  Figure \ref{fig:DecayPressureDependence} plots the measurements for two well-depths; for each well-depth, the dependence of the injected ions (Ne$^{10+}$) only and the sum of detected ions are presented. In the observation range below $\sim\!\! 10^{-6}$ Pa, all observed decay rates are proportional to the gas pressure.  For the base pressure of $1.2 \times 10^{-7}$ Pa obtained in this room-temperature apparatus, a decay time constant of 3.8 s was observed for the sum of detected ions coming from a $\Delta \rm{V} = 10$ V well.  Storage times $\geq 1$ s are useful for a variety of experiments. From the trend in Fig. \ref{fig:DecayPressureDependence}, we anticipate further improvements at lower pressures, e.g. in a cryogenic apparatus. Tests indicate that the sintered NdFeB magnets can be thermal-cycled between liquid nitrogen and room temperatures without damage or degradation.  In addition to lowering the residual gas pressure via cryopumping, operating at 77K would strengthen the remnant field of the NdFeB magnets by $\approx~15$ percent.  With the trap electrodes and magnets heat-sunk to a cryogenic bath, thermal fluctuations can be expected to be significantly smaller. However, the best performance in the low pressure regime would require good surface uniformity  and conductivity to minimize patch effects which contribute to the ion cloud expansion rate.  For the experiments discussed here, all surfaces of magnets and iron electrodes that bound the trapping volume are polished to near optical flatness.  The electrical conductivity of the surfaces can be improved by deposition of a thin layer of gold or carbon (such as Aquadag or Aerodag)\cite{disclaimer}.

Ion temperature is important also.  Kinetic/transport theories of an ion swarm moving in a neutral gas indicate that ion mobility\cite{MobilityIon1973} improves with lower temperature.\cite{MobilityIon1978} This is consistent with our observation that ion cloud radius expansion tends to be slower for colder ions.  Fig. \ref{fig:DecayPressureDependence} shows a significant variation in the decay rate of the total number of detected ions depending upon the trapping well depth: The decay rate in a shallower trapping well ($\Delta \rm{V} = 10$ V) is a factor of $\approx 2$ lower than the decay rate for $\Delta \rm{V} = 40$ V.  To correlate this with temperature, we note that the TOF peaks for the $\Delta \rm{V} = 10$ V well are narrower than corresponding peaks in a $\Delta \rm{V} = 40$ V well, by almost a factor of 2, indicating that the ions stored in the shallower $\Delta \rm{V} = 10$ V well are colder.  The expected higher ion mobility could account for a slower ion cloud expansion and consequently a lower decay rate of the ion count from the $\Delta \rm{V} = 10$ V well.

\section{Summary}
\label{sec:Summary}

 Penning traps with unitary architecture are presented which fully integrate rare-earth magnets and electrical conductors to jointly provide the electric and magnetic fields necessary for ion confinement.  In the one-magnet Penning trap, the simplest case, a Ni-Cu-Ni-plated NdFeB magnet serves as the ring electrode.  In the two-magnet Penning trap, occupying as little as  125 $\rm{cm}^3$ of space, the rare-earth magnets are secluded and yoked by soft-iron electrodes; the reentrant endcaps become the dominant source of the magnetic field in the trapping region. The magnetic field at the center is about 0.32 T with better than 1 percent uniformity over a distance of $\pm 2$ mm.  The simplicity of such compact Penning traps contributes to ease of assembly, self-alignment, operation at high voltages, and beam access to the trapping region.  

We have demonstrated the use of a one-magnet Penning trap and a two-magnet Penning trap for storing highly charged ions extracted from the NIST EBIT.  Ions captured in these room-temperature traps include Ne$^{10+}$, Ne$^{9+}$, Ne$^{8+}$, as well as Ar$^{16+}$, Ar$^{15+}$, Ar$^{14+}$  and Ar$^{13+}$. Experiments with captured bare Ne nuclei indicate that the one-magnet and two-magnet Penning traps are well aligned; the stored ion clouds in both traps expand slowly, with comparable ion counts and loss rates dominated by collisions with background gas.  In initial studies with the one-magnet Penning trap, the ion count rate for short storage time was so high as to exceed the limit of the position-sensitive MCP detector, which exhibits pile-up effects when the ion count rate exceeds $\approx 100$ kHz.  Using data for ion storage times $> 0.5$ second to avoid pile-up effects in the detector for the one-magnet trap, we obtained ion cloud expansion time-constants of order 1 second at base pressure $2.4 \times 10^{-7}$ Pa.  Subsequently, a much faster time-of-flight MCP detector was added in the modified set-up for the two-magnet Penning trap to allow observations in the short storage-time regime, analysis of the ion charge states, and better fine-tuning of ion injection.  Another improvement was the use of more vacuum pumps (not shown in Fig.\ref{fig:Apparatus}) to obtain lower residual background gas pressure.  A base pressure of $1.2 \times 10^{-7}$ Pa was attained in the two-magnet Penning trap; based upon observed decay rates, we estimate that the ion cloud expansion time-constant is about 3 s under optimal conditions, sufficiently long for a variety of spectroscopic experiments. Improvement in storage lifetime is anticipated at lower pressures, e.g. in a cryogenic apparatus.  Active stabilization of the ion cloud rotation to prolong ion confinement is also possible by incorporating ``rotating-wall" electrodes.\cite{RotateWall1997,RotateWall2000} 

Our work is motivated by efforts at NIST to produce one-electron ions in circular Rydberg states.\cite{HCI2010Tan,1eRydberg2008,1eRydberg2009,muonicH2011UDJ2} Other potential applications include portable mass spectrometers (for space-borne or earth-based field instruments), laser spectroscopy, and studies of Rydberg or metastable states in stored highly charged ions.

ACKNOWLEDGMENTS.  The work of Nicholas D. Guise is supported by a National Research Council Associateship Award.


\begin{thebibliography}{41}
\bibitem[{dis()}]{disclaimer}
\bibinfo{note}{Identification of a product herein is for documentation purposes
  only, and does not imply recommendation or endorsement by NIST, nor does it
  imply that this product is necessarily the best available for the purpose.}
\end{thebibliography}

\end{document}